\documentclass[12pt]{iopart}
\usepackage[utf8]{inputenc}
\usepackage{graphicx}
\usepackage{palatino}
\usepackage[pdftex]{color}

\usepackage{bm}
\DeclareBoldMathCommand{\matR}{{\sf R}}
\DeclareBoldMathCommand{\matT}{{\sf T}}
\DeclareBoldMathCommand{\matA}{{\sf A}}
\DeclareBoldMathCommand{\matD}{{\sf D}}
\DeclareBoldMathCommand{\matI}{{\sf I}}
\DeclareBoldMathCommand{\matsig}{{\sf \sigma}}

\DeclareMathAlphabet\mathbfcal{OMS}{cmsy}{b}{n}

\begin{document} 


\title{Radiative heat transfer in 2D Dirac materials}

\author{
Pablo Rodriguez-L\'{o}pez$^{1}$,
Wang-Kong Tse$^{2}$, and
Diego A. R. Dalvit$^{2}$}
\address{$^{1}$ Laboratoire de Physique Th\'{e}orique et Mod\`{e}les Statistiques, CNRS UMR 8626, B\^{a}t. 100, Universit\'{e} Paris-Sud, 91405 Orsay cedex, France}
\address{$^{2}$ Theoretical Division MS B213, Los Alamos National Laboratory, Los Alamos, New Mexico 87545, USA}

\begin{abstract}
We compute the radiative heat transfer between two sheets of 2D Dirac materials, including topological Chern insulators and graphene, within the framework of the local approximation for the optical response of these materials. In this approximation, which neglects spatial dispersion, we derive both numerically and analytically the short-distance asymptotic of the near-field heat transfer in these systems, and show that it scales as the inverse of the distance between the two sheets. Finally, we discuss the limitations to the validity of this scaling law imposed
by spatial dispersion in 2D Dirac materials.
\end{abstract}





\section{Introduction}

Two-dimensional Dirac materials constitute a class of materials 
that carry electrons governed by the Dirac equation. Since the discovery of the first Dirac material graphene \cite{Graphene_RMP}, the past ten years 
have seen another expanding family of such materials now including topological Chern insulators, transition-metal dichalcogenides (TDMC), as well as silicene and germanene \cite{DM_review}. Because of their atomic thickness and unusual electronic properties compared to conventional materials, optical transport properties have received particular attention recently generating prospects for using Dirac materials in graphene photonics \cite{graph_ph}, graphene plasmonics \cite{graph_pl}, and  TDMC-based valleytronics \cite{valley}.  One intriguing feature in the optical properties of 2D gapless Dirac materials is the emergence of a universal optical conductivity, depending only on the fine structure constant that arises from their gapless linear electronic dispersion. Their magneto-optical properties have also been extensively investigated recently. 
In the presence of strong magnetic fields, it was shown \cite{Tse2010,Tse2011} that the magneto-optical Faraday and Kerr effects exhibit quantization and large rotations in the low-frequency regime.  The interesting optical and electronic properties of Dirac materials have spawned a number of works examining the problem of Casimir forces in graphene systems (see, for example, \cite{Bordag2006,Santos2009,Drosdoff2010,Banishev2013}.) Remarkably, a strong magnetic field applied in the out-of-plane orientation between two graphene sheets was shown to result in a quantized Casimir force  \cite{Tse2012}, and a similar effect between Chern insulators without an external magnetic field has also been predicted \cite{Rodriguez-Lopez2014}.

The problem of  radiative heat transfer \cite{Polder1971,Joulain2005,Volokitin2007} is intimately related to the Casimir effect  since they both are examples of
dispersion interactions, and they respectively originate from the exchange of energy and momentum between two bodies mediated by fluctuating electromagnetic
fields.  Both phenomena can be described in terms of the optical transmission and reflection properties of the cavity formed by the two bodies.  
Near field heat transfer has been recently measured in bulk materials \cite{Kittel2005,Narayanaswamy2008,Rousseau2009,Ottens2011,Kralik2012}.
Previous theory works on radiative heat transfer in graphene have focused on the metallic regime where the Fermi level lies in one of the bands. The near-field heat transfer is dominated by 
the effects of surface plasmon polaritons corresponding to the optical Drude response near the Fermi level \cite{Svetovoy2012,Ilic2012}. 
Recently near-field heat transfer between graphene-coated bulk materials has been measured \cite{Zwol2012b}, showing that thermally excited graphene plasmons can substantially increase radiative heat transfer at short separations. Lifting of the degenerate Dirac point by time-reversal symmetry (as in the case of Chern insulators) or by spatial inversion symmetry (as in the case of 2D TDMC) results in a Dirac insulator with a band gap $\Delta$. When the chemical potential lies within
the band gap, the plasmonic contribution to the radiative heat transfer is exponentially suppressed and the heat transfer is mainly due to interband optical transitions. 

The short-distance asymptotics of dispersion interactions depends on the dimensionality and optical  properties of the involved bodies, and has been studied in the past
for selected materials. For example, the van der Waals/Casimir  interaction energy between planar bodies at thermal equilibrium has been computed using van der Waals energy functionals  \cite{Dobson2006}, and scales as $1/L^2$ for 3D metals and insulators, as $1/L^4$ for 2D insulators, and as $1/L^{5/2}$ for 2D metals. In the case of thermal non-equilibrium, the corresponding van der Waals energy for 3D metals and insulators has been shown to retain the same $1/L^2$ scaling law as in the thermal equilibrium case
\cite{Bimonte2011}. Near-field heat transfer is another manifestation of dispersion interactions in thermal non-equilibrium, and its short-distance asymptotics is known for various 3D systems. 
In the regime when spatial dispersion can be neglected and the optical conductivity is independent of momentum, near field heat 
transfer scales as $\simeq L^0$ for good conductors, and there is a crossover  to a $1/L^2$ dependency for high resistivity metals  \cite{Volokitin2007,Volokitin2001}. In addition, a thin (quasi-2D) high-resistivity coating of a 3D good metal can drastically increase the heat transfer between two solids, and modify the $L^0$-asymptotics to the $1/L^2$ law \cite{Volokitin2007,Volokitin2001}. This same $1/L^2$ asymptotics is obtained for polar materials (e.g, SiC) \cite{Joulain2005}. 
However, all these scaling laws predict an unphysical divergence at very short separations. In this regime, non-local optical response (spatial dispersion) becomes relevant, strongly modifying how near-field heat transfer varies with distance. In the case of  metals \cite{Chapuis2008}, effects of spatial dispersion in radiative heat transfer show up for distances below 0.1 nm between two parallel surfaces and result in the modification of the $1/L^2$ law, the amount of heat transfer saturating at short separations. Similar effects occur for dielectrics \cite{Singer2015}, although the deviation from the $1/L^2$ law occurs at larger distances of the order of a few nanometers. In both cases, at even shorter separations quantum effects become relevant, and the whole description of radiative transfer based on macroscopic electrodynamics breaks down. Finally, it is worth stressing that, for given materials,  the distance regime where spatial dispersion effects start to show up depend on the observable in question, e.g. heat transfer, quantum friction, etc.

Here, we investigate the radiative heat transfer between two-dimensional Dirac materials within the local approximation for the optical response. In particular, we study the asymptotic short-distance behavior of near-field heat transfer in these systems, including topological Chern insulators and graphene, and show a $1/L$ scaling law at short separations. As in the case of 3D metals and dielectrics mentioned above, we expect this scaling law to break down at very short separations. To the best of our knowledge, there is no widely accepted treatment  in the literature of nonlocal effects on the optical response of 2D Dirac materials, which prevents us from assessing the precise region of validity of the local approximation used in this work. We briefly discuss this at the end of the paper.


\section{Optical response of 2D Dirac materials and radiative heat transfer}

In this section we review the derivation of the Fresnel reflection coefficients for 2D Dirac materials. We assume the 2D material is located at $z=0$ and a plane wave 
is impinging on the 2D surface from the $z>0$ side. The wave-vector ${\bf k}$ and the normal to the interface ${\bf n} = \hat{z}$ define the plane of incidence. We choose
a $x-y$ coordinate system on the surface so that $\hat{x}$ is parallel to the plane of incidence and $\hat{y}$ is orthogonal to it. The electromagnetic response of the 2D film is characterized by a surface current $\mathbfcal{J} = (4 \pi/c) \matsig \cdot \mathbfcal{E}$, where $\mathbfcal{E}$ is the component of the electric field ${\bf E}$ on the $(x,y)$ plane. The optical conductivity tensor is
\begin{eqnarray}
\matsig(\omega) = \left(\begin{array}{c|c}
\sigma_{xx}(\omega) & \sigma_{xy}(\omega) \\
\hline
\sigma_{yx}(\omega) & \sigma_{yy}(\omega)
\end{array}\right)
\end{eqnarray}
For s-polarized (transverse electric) waves, the electric field is in the $y$ direction (orthogonal to the plane of incidence), so that $\sigma_{yy}(\omega)$ is the longitudinal conductivity and $\sigma_{yx}(\omega)$ is the  Hall conductivity. For p-polarized (transverse magnetic) waves, the electric field is 
in the $x-z$ incidence plane, so that $\sigma_{xx}(\omega)$ is the longitudinal conductivity and $\sigma_{xy}(\omega)$ is the  Hall conductivity.
Here and in the following we assume that one can neglect any momentum dependency of the conductivity, i.e. we discard any effects of spatial dispersion.
We further assume that the 2D material is isotropic, so that $\sigma_{yy}(\omega)= \sigma_{xx}(\omega)$ and
$\sigma_{yx}(\omega)= - \sigma_{xy}(\omega)$. Imposing the boundary conditions for the electromagnetic field, ${\bf n} \times {\bf H} = \mathbfcal{J}$ and ${\bf n} \times {\bf E} =0$, one can obtain the Fresnel reflection and transmisson matrices of the 2D material. In particular, the components of the $2 \times 2$ reflection matrix $\matR$ have the form
\cite{Tse2011,Rodriguez-Lopez2014,Grushin2011}
\begin{eqnarray}
r_{ss}(\omega,{\bf K}) & = & - \frac{2  \pi}{\cal D}  \left[ \frac{\sigma_{xx}(\omega)}{c \xi} + \frac{2 \pi}{c^2} (\sigma_{xx}^2(\omega) + \sigma_{xy}^2(\omega)) \right]  \\
r_{sp}(\omega, {\bf K}) & = &  r_{ps}(\omega,{\bf K})  = \frac{2  \pi}{\cal D}  \frac{1}{c} \sigma_{xy}(\omega) \\
r_{pp}(\omega,{\bf K}) &=& \frac{2 \pi}{\cal D} \left[ \frac{\sigma_{xx}(\omega)}{c}  \xi+ \frac{2 \pi}{c^2} ( \sigma_{xx}^2(\omega) + \sigma_{xy}^2(\omega) ) \right] 
\label{reflection-coeff}
\end{eqnarray}
Here  ${\bf K}$ is the component of the wave vector ${\bf k}$ on the $(x,y)$ plane, $\xi=k_z c /\omega$, ${\cal D} = 1 + 2 \pi \frac{\sigma_{xx}(\omega)}{c} \left( \frac{1}{\xi} + \xi \right) + \frac{4 \pi^2}{c^2} (\sigma_{xx}^2(\omega) + \sigma_{xy}^2(\omega))$, $k_z = \sqrt{\omega^{2}/c^{2} - K^{2}}$ is the photon momentum normal to the interface ($K=|{\bf K}|$) and $\omega$ is the frequency.

The complex longitudinal $\sigma_{xx}(\omega) = \sigma_{xx,R}(\omega) + i \sigma_{xx,I}(\omega)$ and Hall 
$\sigma_{xy}(\omega) = \sigma_{xy,R}(\omega) + i \sigma_{xy,I}(\omega)$ conductivities for 2D Dirac materials have been computed
using different approaches, e.g. the Kubo formula \cite{Rodriguez-Lopez2014} and the quantum kinetic equation \cite{Tse2010}. In the limit of low temperatures $k_B T  \ll {\rm min}(|\mu_F|, |\Delta|)$ (here $\mu_F$ is the Fermi energy relative to the Dirac point) and for small disorder, the dissipative components of the optical conductivity are given by \cite{Tse2010,Tse2011}
\begin{eqnarray}
\frac{\sigma_{xx,R}(\omega)}{\alpha c} &=& \theta(|\mu_F| - |\Delta|) \delta(\hbar \omega) \frac{\mu_F^2-\Delta^2}{4 |\mu_F|} +
\left( \frac{1}{16} + \frac{\Delta^2}{4 \hbar^2 \omega^2} \right) \theta[\hbar |\omega| - 2 \; {\rm max}(|\mu_F|, |\Delta|)] \nonumber \\
\frac{\sigma_{xy,I}(\omega)}{\alpha c} &=& \frac{\Delta}{4 \hbar \omega} \theta[\hbar |\omega| - 2 \; {\rm max}(|\mu_F|, |\Delta|)]
\label{sigmas-DIS}
\end{eqnarray}
where $\alpha=e^2/\hbar c = 1/137$ is the vacuum fine structure constant, and $\theta(x)$ is the Heaviside step function. 
The reactive components, which are due to off-shell virtual transitions, are given by
\begin{eqnarray}
\frac{\sigma_{xx,I}(\omega)}{\alpha c} &=& \theta(|\mu_F| - |\Delta|) \frac{\mu_F^2-\Delta^2}{4 \pi \hbar \omega |\mu_F|}  \nonumber \\
&& + \frac{1}{16\pi} \left[ \frac{4 \Delta^2}{\hbar \omega} ( [{\rm max}(|\mu_F|,|\Delta|)]^{-1} - \epsilon_c^{-1} ) +
\left(1+ \frac{4 \Delta^2}{\hbar^2 \omega^2} \right) f(\omega) \right] \nonumber \\
\frac{\sigma_{xy,R}(\omega)}{\alpha c} &=&- \frac{\Delta}{4 \pi \hbar \omega} f(\omega)
\label{sigmas-REA}
\end{eqnarray}
where $f(\omega)= \ln |(\hbar \omega+2 \epsilon_c)/(\hbar \omega-2\epsilon_c)| - \ln |[\hbar \omega - 2 \; {\rm max}(|\mu_F|,|\Delta|)]/[\hbar \omega-2 \; {\rm max}(|\mu_F|,|\Delta|)]|$, and
$\epsilon_c$ is the energy cutoff of the Dirac Hamiltonian, which we associate with the separation between the Dirac point and the closest bulk band.
In the rest of this paper we will neglect small quantitative corrections from the finite energy cutoff and take $\epsilon_c \rightarrow \infty$.
The above expressions for the reflection properties can be readily used to compute the near-field radiative heat transfer between 2D Dirac materials (or, for example, between
a 2D Dirac material and a metal or a dielectric). 

Two vacuum-separated bodies held at different temperatures $T_1>T_2$ exchange energy through radiative heat transfer \cite{Polder1971}. The heat transfer per unit area between two 
semi-infinite parallel plates separated by a distance $L$ can be computed from the $z$ component  (normal to the planes) of the Poynting vector,
\begin{equation}
\langle S_z \rangle = \int_{0}^{\infty}\frac{d\omega}{2\pi} \hbar \omega \left[  n_B(\omega, T_{1}) - n_B(\omega, T_{2}) \right] 
\int\frac{d^{2} {\bf K}}{(2\pi)^{2}} \mathcal{T}(\omega, \textbf{K})
\label{heat-transfer}
\end{equation}
where $n_B(\omega, T) =  [\exp(\hbar \omega / k_{\rm B}T)-1]^{-1}$
is the mean occupation of a harmonic oscillator of frequency $\omega$ at temperature $T$, and $\mathcal{T}(\omega, \textbf{K})$ is the transfer factor defined as
\begin{eqnarray}
\mathcal{T}(\omega, \textbf{K}) = \left\lbrace \begin{array}{ll}
{\rm Tr} \left[  \left( \matI - \matR_{2}^{\dagger} \matR_{2} \right) \matD  \left( \matI - \matR_{1}^{\dagger} \matR_{1} \right) \matD^{\dagger} \right]  & {\rm for} \;  K < \frac{\omega}{c}\\
{\rm Tr} \left[ \left( \matR_{2} - \matR_{2}^{\dagger} \right) \matD \left( \matR_{1}^{\dagger} - \matR_{1} \right) \matD^{\dagger} \right] 
e^{-2 L  |k_z |}  &  {\rm for} \; K > \frac{\omega}{c}\\
\end{array}\right.
\end{eqnarray}
Here $\matI$ is the $2\times2$ identity matrix and $\matD= \left( \matI - \matR_{1} \matR_{2} e^{2i\,k_{z} L} \right)^{-1}$ is a Fabry-Perot-like denominator describing the effect of multiple reflections.
Values of $K < \omega/c$ correspond to propagative electromagnetic waves in vacuum, and $K > \omega/c$ correspond to evanescent waves in vacuum. Both of these waves contribute to the radiative heat transfer at a given distance $L$. In the near-field, the heat exchange can be enhanced by several orders of magnitude due to tunneling of electromagnetic evanescent waves between the two bodies. Furthermore, this enhancement can be made even stronger for certain materials that are able to support surface modes in the IR part of the electromagnetic spectrum, such as surface phonon polaritons in polar materials \cite{Joulain2005} or plasmon polaritons in nanostructured metals \cite{Guerout2012}. We note that when the optical properties of the involved materials depend on temperature, the transfer factor is temperature dependent, 
$\mathcal{T}(\omega, {\bf K}; T_1, T_2)$. 

In the case of parallel plane slabs of {\it finite thickness} (and also for parallel 2D sheets), the expression of the transfer factor in the 
evanescent sector remains unchanged, while in the propagative sector ($K < \omega/c$) it is modified as \cite{Krueger2012}
\begin{equation}
\mathcal{T}_{\rm slabs}(\omega, \textbf{K}) = {\rm Tr} \left[  \left( \matI - \matR_{2}^{\dagger} \matR_{2} -\matT_{2}^{\dagger} \matT_{2}
\right) \matD  \left( \matI - \matR_{1}^{\dagger} \matR_{1} - \matT_{1}^{\dagger} \matT_{1} \right) \matD^{\dagger} \right] ,
\end{equation}
where $\matT$ is the $2 \times 2$ transmission matrix of a slab. In the case of 2D materials, the components of the transmission matrix are given by
\begin{eqnarray}
t_{ss}(\omega,{\bf K}) & = &  \frac{1}{\cal D} \left[ 1 + 2 \pi \xi \frac{\sigma_{xx}(\omega)}{c} \right]  \\
t_{sp}(\omega,{\bf K}) & = & - t_{ps}(\omega,{\bf K}) = - \frac{2\pi}{\cal D}  \frac{1}{c} \sigma_{xy}(\omega)\\
t_{pp}(\omega,{\bf K}) & = & \frac{1}{\cal D}  \left[ 1 + 2\pi\frac{\sigma_{xx}(\omega)}{c \xi} \right]
\label{transmision-coeff}
\end{eqnarray}
where $\cal D$ and $\xi$ have been defined below Eq.~\ref{reflection-coeff}.


\section{Radiative heat transfer for 2D Chern insulators}

The optical conductivity tensor of a 2D Chern insulator can be obtained from (\ref{sigmas-DIS},\ref{sigmas-REA}) for the case $|\mu_F| < |\Delta|$. In the following we take
$\mu_F=0$ and $\Delta \neq 0$. We have
\begin{eqnarray}
\frac{\sigma_{xx}(\omega)}{\alpha c} &=& 
\left( \frac{1}{16} + \frac{\Delta^{2}}{4 \hbar^2 \omega^{2}} \right) \theta\left( \hbar \omega - 2 |\Delta| \right)  + 
i \left[ \frac{|\Delta|}{4\pi \hbar \omega } - \frac{1}{16\pi} \left(1 + \frac{4\Delta^{2}}{\hbar^2 \omega^{2}}\right) \log \left| \frac{ \hbar \omega + 2 |\Delta| }{ \hbar \omega - 2 |\Delta|} \right| \right] \nonumber\\
\frac{\sigma_{xy}(\omega)}{\alpha c} &=& \frac{\Delta}{4 \pi \hbar \omega } \log \left| \frac{\hbar \omega + 2 |\Delta|}{ \hbar \omega - 2 |\Delta|} \right| +
i \frac{\Delta}{4 \hbar \omega}\theta\left( \hbar \omega - 2 |\Delta| \right)
\label{sigmas-CI}
\end{eqnarray}
We first study analytically the far- and near-field regimes of the radiative heat transfer between 2D Chern insulators.


\begin{figure}[t] 
\centering
\includegraphics[height=8cm]{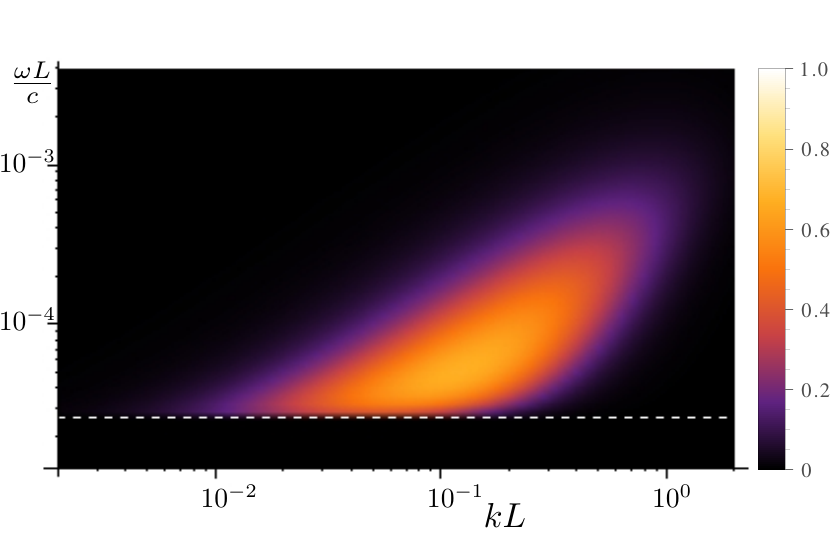} 
\caption{(Color online) Transfer factor $\mathcal{T}(\omega,{\bf K})$ for two identical 2D Chern insulators in the near-field regime.
We plot only the evanescent sector ($K>\omega/c$). 
Parameters are  $L=1$  nm, $T_1=10$ K, $T_2=1$ K, and $\Delta/k_B=30$ K. The horizontal dashed line corresponds to $\omega L/c =2 \Delta L/(\hbar c) =1.4 \times 10^{-5}$, below which the transfer factor identically vanishes. Contributions of frequencies above $\omega L/c= k_B L \; {\rm max}(T_1, T_2)/(\hbar c)=4.6 \times 10^{-6}$  are exponentially suppressed due to the Bose factors in (\ref{heat-transfer}).}
\label{fig:Transmission-CI}
\end{figure}

\begin{figure}[t] 
\centering
\includegraphics[height=7cm]{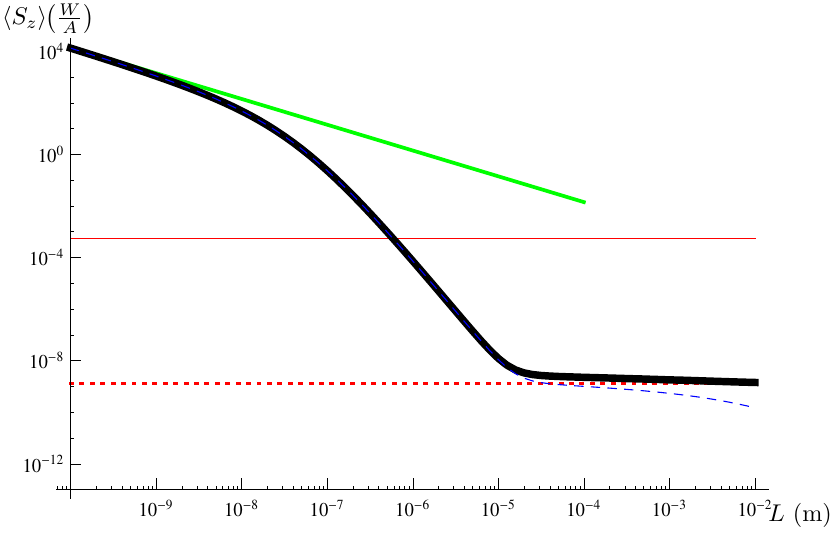} 
\caption{(Color online) Radiative heat transfer between two identical 2D Chern insulators as a function of their separation within the local approximation for the optical response. The solid black curve is the numerical evaluation of Eq. (\ref{heat-transfer}), the dashed blue curve is the numerical evaluation of the evanescent contribution only, and the dotted red curve is the numerical evaluation of the propagative contribution only. Note that the far-field heat transfer is several orders of magnitude smaller than for two black bodies, given by the Stefan-Boltzmann law (horizontal thin red line). The thin solid green curve corresponds to the analytical expression  (\ref{analytics-CI}) that gives the $1/L$ small distance asymptotics. Parameters are $T_1=10$ K, $T_2=1$ K, and $\Delta/k_B=30$ K.}
\label{fig:HT-CI}
\end{figure}


In the far-field regime, $L \gg \lambda_{\rm th}$ (here $\lambda_{\rm th}=\hbar c/ k_{\rm B} T$ is the thermal wavelength), the evanescent field of one body cannot reach the other one, and therefore only propagative modes contribute to the heat transfer in this case. In addition, since the 2D Dirac materials are very thin with conductivities proportional to $\alpha \ll 1$, their reflection matrices are small, and one can approximate $\matD \approx \matI$. Therefore,
$\mathcal{T}(\omega, \textbf{K}) \approx \theta(\omega/c -K) {\rm Tr}[ (\matA^{\dagger}_2 \matA_2) (\matA^{\dagger}_1 \matA_1)]$,
where $\matA^{\dagger} \matA = \matI - \matR^{\dagger} \matR - \matT^{\dagger} \matT $ is the absorptivity of the 2D sheet, which depends both on frequency and momentum through the reflection and transmission matrices. Due to their weak coupling to light, the absorptivity of 2D Dirac materials is typically very small (e.g., for a suspended graphene sheet it is given by $\pi e^2/\hbar c = 2.3 \%$ at normal incidence and for $k_B T \ll \mu_F < \hbar \omega/2$), the far-field radiative heat transfer is much smaller than that for two perfect black bodies (perfect absorbers/emitters at all frequencies and momenta), given by $S_{BB} = \sigma (T_1^4 - T_2^4)$, where $\sigma= \pi^2 k_{\rm B}^4/60 \hbar^3 c^2$ is Stefan's constant. 

In the near-field regime, $L \ll \lambda_{\rm th}$, the radiative heat transfer can be substantially enhanced with respect to the far-field value when the evanescent fields of the two bodies hybridize and open new channels for energy transfer. This requires that the spectra of the two bodies are matched in frequency and that are lie within the thermal envelope defined by $\hbar \omega [n_B(\omega, T_{1}) - n_B(\omega, T_{2})]$ so that there is a non-negligible contribution to the $\omega$-integral. In addition, a further increase can take place when the two bodies possess surface modes (surface phonon polaritons in the case of dielectrics, and surface
plasmon polaritons in the case of metals) that are frequency-matched.
In this regime we can neglect the contribution from the propagative modes, and concentrate solely on the evanescent part. 
Because of the Bose factors in Eq.(\ref{heat-transfer}), the most relevant frequencies contributing to the heat exchange are $\omega  \approx k_B T_1/\hbar, k_B T_2/\hbar$, and since the optical response of the 2D Chern insulator given in  (\ref{sigmas-CI}) is valid for $k_B T \ll \Delta$, it implies that the most relevant frequencies are in the range $\omega \ll |\Delta|/\hbar$. In this regime, the dissipative components of the optical conductivity are approximately zero,  
$\sigma_{xx,R}(\omega) \approx 0$ and $\sigma_{xy,I}(\omega) \approx 0$, and the reactive components are given by
$\sigma_{xx,I}(\omega) \approx - \frac{\alpha c}{12 \pi} \frac{\hbar \omega}{\Delta}$ and $\sigma_{xy,R}(\omega) \approx {\rm sgn}(\Delta) \alpha c/4 \pi$. In the evanescent sector, 
$k_{z}= i \sqrt{K^2-\omega^2/c^2} \equiv i \kappa$ is purely imaginary, and then from (\ref{sigmas-CI}) one concludes that the imaginary parts of all the reflection amplitudes, and hence the matrices
$\matR - \matR^{\dagger}$ appearing in (\ref{heat-transfer}), vanish for $\omega < |\Delta|/\hbar$. Therefore, for the most relevant frequencies 
$\omega \approx  k_B T_1/\hbar, k_B T_2/\hbar$ contributing to the evanescent heat transfer, there is no heat transfer whatsoever. Only for frequencies $\omega$ larger than 
$|\Delta|/\hbar$ can one obtain a non-vanishing evanescent contribution to the transfer, although it is exponentially suppressed by the Bose factor as $\exp(- |\Delta| / k_B T)$. 

We now study analytically the short-distance asymptotics of the radiative heat transfer between two identical 2D Chern insulators, that is given by the evanescent sector, as discussed above. It is convenient to introduce dimensionless variables $u= \hbar \omega / 2 |\Delta|$ (this simplifies the form of the optical conductivities ({\ref{sigmas-CI}))
and $t=L \kappa$ (recall that $\kappa=\sqrt{K^2-\omega^2/c^2}$, so $K dK =\kappa d\kappa$ in the integrand of (\ref{heat-transfer})). We then rewrite the expression for the heat transfer as $\langle S_z \rangle= I_{{\rm ev},T_1}(L) - I_{{\rm ev},T_2}(L)$, where
\begin{equation}
I_{{\rm ev},T}(L) = \frac{\Delta^2}{\hbar \pi^2 L^2} \int_0^{\infty}  \frac{u du }{e^{2 u |\Delta| / k_B T} -1} \int_0^{\infty} dt \; t  {\cal T}(u,t)
\label{near-field}
\end{equation}
with 
\begin{equation}
{\cal T}(u,t) =  {\rm Tr} \left[ (\matR_{1} - \matR_{1}^{\dagger}) (\matI-\matR_{1} \matR_{2} e^{-2 t})^{-1}  (\matR_{2}^{\dagger}-\matR_{2})  (\matI-\matR_{1}^{\dagger} \matR_{2}^{\dagger} e^{-2 t})^{-1} \right] \e^{-2 t}
\end{equation}
The reflection matrices $\matR(\omega,{\bf K})$ are evaluated at $\omega=2 |\Delta| u/\hbar$ and $k_z=i t/L$. Note that the $L$-dependency of the integrals in (\ref{near-field}) only arises from the $k_z$ dependency of the reflection coefficients. Performing a Taylor expansion of the integrand $t {\cal T}(u,t)$ in powers of $L$, it turns out that the leading contribution is linear in $L$ and is independent of the Hall conductivity $\sigma_{xy}(u)$. After performing the integral over $t$, 
we obtain
\begin{equation}
I_{{\rm ev},T}(L)  \approx \frac{|\Delta|^3}{2 \pi^3 \hbar^2} \; \frac{1}{L} \int_0^{\infty} \frac{u^2 du}{e^{2 u |\Delta| / k_B T} -1} \; 
\frac{\sigma_{xx,R}(u) \left[ \frac{\pi}{2}  + \tan^{-1}( \frac{\sigma_{xx,I}(u)}{\sigma_{xx,R}(u)}) \right]}
{\sigma_{xx,R}^2(u) + \sigma_{xx,I}^2(u)}
\label{analytics-CI}
\end{equation}
that is, a $1/L$ small-distance asymptotics.

In the following we  numerically evaluate the full expression (\ref{heat-transfer}) using the conductivity tensor for 2D Chern insulators  (\ref{sigmas-CI}).
In Fig. (\ref{fig:Transmission-CI}) we show a density plot of the transfer factor $\mathcal{T}(\omega,{\bf K})$ in the evanescent sector for two identical 2D Chern insulators held at temperatures $T_1=10$ K and $T_2=1$ K as a function of  $\omega  L/c$ and $k L$. We choose a value of the gap equal to $\Delta/k_B=30$ K, which lies within the range $2-3$ meV reported in \cite{Chang2013} for the first experimentally demonstrated Chern insulator made of chromium-doped (Bi,Sb)$_2$Te$_3$. As explained above, 
in the near-field regime, the transfer factor vanishes for frequencies below $2 \Delta/\hbar$ (horizontal dashed line in fig. (\ref{fig:Transmission-CI})), which is above the thermal envelope  determined by the Bose factors in (\ref{heat-transfer}). Therefore, the non-vanishing region in the $(\omega,k)$ space of the transfer factor 
in the near field is outside the thermal envelope and results in an exponentially suppressed radiative heat transfer for short separations. 
In Fig.(\ref{fig:HT-CI})  we show the radiative heat transfer as a function of separation between the two 2D Chern insulators.
The solid black curve is the full numerical evaluation of the radiative heat transfer (\ref{heat-transfer}). 
We observe an excellent agreement with the short distance limit given in 
(\ref{analytics-CI}) (thin solid green curve). 
The evanescent contribution is the dashed blue curve, which has an approximate $1/L^{4}$ dependency in the range of distances $10^{-7} \; {\rm m} < L < 10^{-5} \; {\rm m}$ for the parameters chosen. The propagative contribution is the dotted red curve. In the far-field, the radiative heat transfer between two Chern insulators is several orders of magnitude smaller than that between two perfect black bodies.


\begin{figure}[t] 
\centering
\includegraphics[height=8cm]{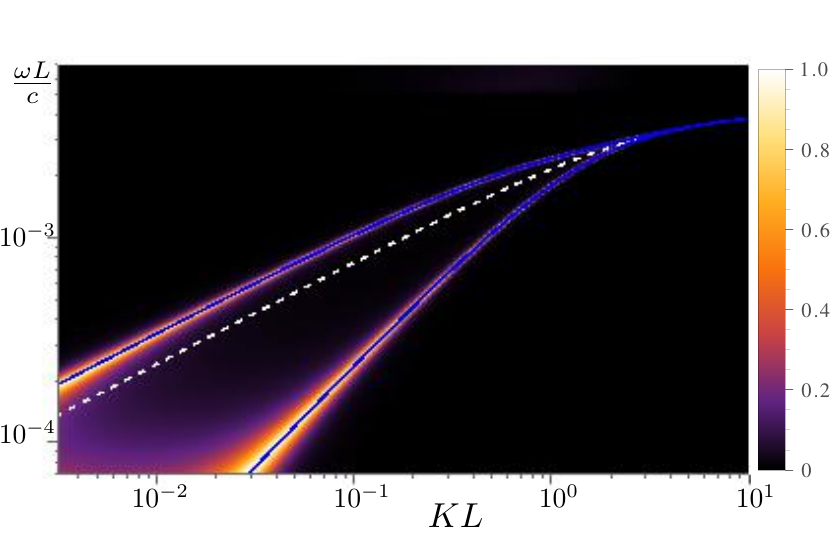}
\caption{(Color online) Transfer factor $\mathcal{T}(\omega,{\bf K})$ for two identical graphene sheets separated by $L=10$ nm. We plot only the evanescent sector ($K>\omega/c$). 
The chemical potential
is $\mu_F=0.5$ eV, $\tau^{-1}=10^{13} \; {\rm s}^{-1}$, and the optical conductivity is computed with Eq.(\ref{correct-graphene}). The evanescent surface plasmon modes of the graphene cavity are evident in the figure. 
The dispersion relation of the plasmonic mode of a single graphene sheet (pole of $r_{pp}(\omega,{\bf K})$) is represented by the dashed white line, and that for the even and odd cavity modes 
(solutions of $1-r^2_{pp}(\omega, {\bf K}) e^{-2 L \sqrt{K^2-\omega^2/c^2}}=0$) are represented by the solid blue lines.}
\label{fig:Transmission-graphene}
\end{figure}


\begin{figure}[t] 
\centering
\includegraphics[height=7cm]{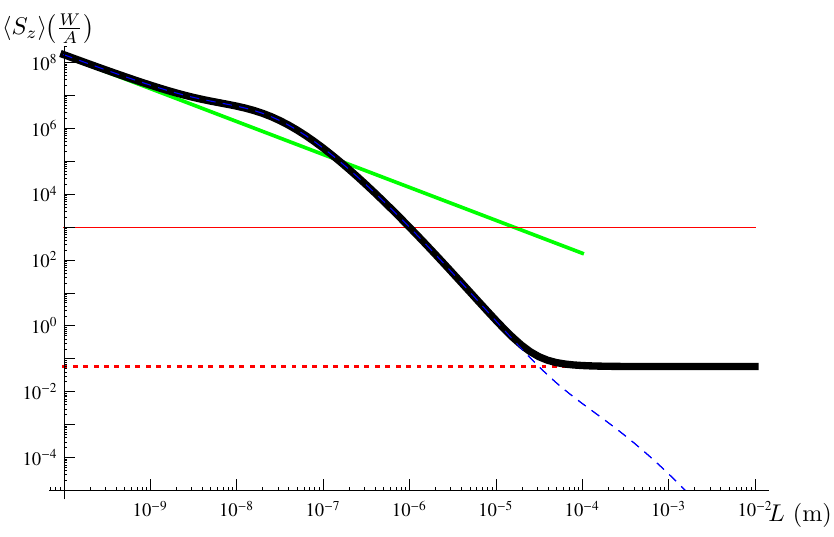} 
\caption{(Color online) Radiative heat transfer between two identical graphene sheets as a function of their separation, within the local approximation for the optical response. The solid black curve is the numerical evaluation (\ref{heat-transfer}), the dotted red curve is the propagative contribution only, the dashed blue curve is the evanescent contribution only, and the thin solid green curve corresponds to the analytical expression  (\ref{analytics-graphene}) that gives the $1/L$ small distance asymptotics. The horizontal thin red line is the Stefan-Boltzmann heat transfer between perfect black bodies.
Parameters are $T_1=400$ K, $T_2=300$ K, $\mu_F=0.5$ eV, and
$\tau^{-1}=10^{13} \; {\rm s}^{-1}$.}
\label{fig:HT-graphene}
\end{figure}


\section{Near-field asymptotics for radiative heat transfer in graphene}

As mentioned in the Introduction, several authors have studied near-field radiative heat transfer in graphene systems {\cite{Svetovoy2012,Zwol2012b,Ilic2012,Drosdoff2014},
which is an interesting platform due to the possibility of tuning the heat exchange via chemical doping or electromagnetic gating. Graphene is known to support
surface plasmon modes in the terahertz to near-infrared part of the spectrum, and these modes are responsible for a strong enhancement of the near-field
heat transfer at room temperatures. In this section we will focus on the near-field asymptotics of heat transfer in graphene, and derive how it scales
with the distance $L$.

For the case of graphene, a gapless  ($\Delta=0$) 2D Dirac material, it follows from (\ref{sigmas-DIS}) and (\ref{sigmas-REA}) that
the Hall conductivity is identically zero, $\sigma_{xy}(\omega)=0$, and the longitudinal conductivity is given by (we assume $\mu_F>0$)
\begin{equation}
\frac{\sigma_{xx}(\omega)}{\alpha c} = 
\left[ \mu_F \delta(\hbar \omega) + \frac{1}{4} \theta(\hbar \omega-2 \mu_F) \right] + i
\left[ \frac{\mu_F}{\pi \hbar \omega} + \frac{1}{4\pi} \log \left| \frac{\hbar \omega - 2 \mu_F}{ \hbar \omega + 2 \mu_F} \right|  \right]
\label{graphene-cond}
\end{equation}
where we have taken into account that graphene possesses 2 Dirac cones each having two-fold spin degeneracy. This expression for the conductivity is valid as long as 
$k_B T \ll \mu_F$ and disorder is negligible. The essential contribution from plasmon excitations is captured by the Drude term,
$\mu_F \delta(\hbar \omega) + i \mu_F / (\pi \hbar \omega)$. 
We note that for graphene the conductivity tensor $\matsig(\omega)$ and the reflection matrix $\matR(\omega,{\bf K})$ are diagonal, and therefore the heat transfer (\ref{heat-transfer}) can be expressed as a sum of the TE (s) and TM (p) polarizations. As shown in previous works \cite{Svetovoy2012,Zwol2012b,Ilic2012}, the TM polarization dominates the near-field heat transfer in graphene. 

Other expressions for graphene's conductivity, including temperature corrections, have been used in the past for modeling radiative heat transfer 
in graphene systems. At finite temperatures,  
the intraband (Drude) and interband components of the conductivity $\sigma^{\rm F}(\omega) = \sigma_D  + \sigma_I$ can be written respectively as  \cite{Falovsky2007}
\begin{eqnarray}
\sigma_D(\omega) &=& \frac{2 i \alpha c k_B T }{\omega + i \tau^{-1}}  \ln[ 2 \cosh(\mu_F/2 k_B T)]  \nonumber \\
\sigma_I(\omega) &=& \frac{\alpha c}{4} 
\left[ 
G(\hbar \omega/2) + i \frac{4 \hbar \omega}{\pi} \int_0^{\infty}  d\xi \frac{G(\xi) - G(\hbar \omega/2)}{(\hbar \omega)^2 - 4 \xi^2}
\right]
\label{falk}
\end{eqnarray}
where $G(\xi) = \sinh(\xi/k_B T) / [\cosh(\mu_F/k_B T) + \cosh(\xi/k_B T)]$. Here, the delta function has been explicitly modeled as 
a Lorentzian function to capture disorder effects. We note, however, that in equation (\ref{falk}), disorder broadening was only included in the Drude contribution but was neglected in the interband contribution \cite{Falovsky2007}. 
In the limit of low temperatures ($k_B T \ll \mu_F$), equation (\ref{falk}) takes the form 
 \begin{equation}
 \frac{\sigma^{\rm F}(\omega)}{\alpha c} = 
 \left[ \frac{\mu_F}{\pi \hbar} \frac{\tau^{-1}}{\omega^2+\tau^{-2}} + \frac{1}{4} \theta(\hbar \omega-2 \mu_F) \right] + i
 \left[ \frac{\mu_F}{\pi \hbar} \frac{\omega}{\omega^2+\tau^{-2}} + \frac{1}{4\pi} \log \left| \frac{\hbar \omega - 2 \mu_F}{ \hbar \omega + 2 \mu_F} \right|  \right]
 \label{correct-graphene}
 \end{equation}
which is consistent with equation (\ref{graphene-cond}) for $\tau^{-1} \ll \omega$. 
For example, for a Fermi energy $\mu_F = 0.5 $ eV (or $5800$ K, corresponding to an electron density of 
$n = \frac{\mu_F^2}{\pi \hbar^2 v^2} \approx 10^{13} \; {\rm cm }^{-2}$ for a velocity parameter of $v = 10^8$ cm/s), $\tau$ of the order of $10^{-13}$ s, 
and for temperatures $T<1000$ K, the expression (\ref{correct-graphene}) is an excellent description to the temperature-dependent conductivity
(\ref{falk}), and also agrees with (\ref{graphene-cond}) for $\omega\gg \tau^{-1}$.
In the following we will  use the low-temperature formula (\ref{correct-graphene}) for the conductivity of graphene. 
We show in Fig.(\ref{fig:Transmission-graphene}) the transfer factor in the evanescent sector for two identical graphene sheets separated by a distance $L=10$ nm. The obtained transfer factor is identical to the one shown in
Fig. 2 of \cite{Ilic2012} that was computed with the finite-temperature conductivity (\ref{falk}) for two identical graphene sheets with $\mu_F=0.5$ eV,
$\tau=10^{-13}$ s,  and held at temperature $T=300$ K.

We now study the behavior of heat transfer between two identical graphene sheets as a function of separation. In the far-field limit, due to the small absorptivity of graphene, we obtain that the heat transfer becomes independent of distance (as in the Stefan-Boltzmann law), but several orders of magnitude lower.
In the near-field regime, we obtain the following approximate analytical expression for the asymptotic short-distance behavior
\begin{equation}
I_{{\rm ev},T}(L)  \approx \frac{\mu_F^3}{2 \pi^3 \hbar^2} \; \frac{1}{L} \int_0^{\infty} \frac{u^2 du}{e^{2 u \mu_F / k_B T} -1} \; 
\frac{\sigma_{xx,R}(u) \left[ \frac{\pi}{2}  + \tan^{-1}( \frac{\sigma_{xx,I}(u)}{\sigma_{xx,R}(u)}) \right]}
{\sigma_{xx,R}^2(u) + \sigma_{xx,I}^2(u)}
\label{analytics-graphene}
\end{equation}
i.e., a $1/L$ asymptotics at short separation. Note that this expression is identical to that for Chern insulators (\ref{analytics-CI}) with the replacement
$\Delta \rightarrow \mu_F$.
Finally, we show in Fig.(\ref{fig:HT-graphene})  the radiative heat transfer as a function of separation between two identical graphene sheets held at temperatures $T_1=400$ K and $T_2=300$ K. 
The solid black curve is the full numerical evaluation of the radiative heat transfer, the dashed blue curve is the evanescent contribution only (which has an approximate $1/L^{3}$ dependency at the larger separations shown), and the thin solid line is the $1/L$
short distance asymptotic given by (\ref{analytics-graphene}).


\section{Conclusion}

We have studied radiative heat transfer between two 2D Dirac materials scales using the local approximation for the optical response of these materials, i.e. neglecting effects of spatial dispersion, and derived analytical expressions for the near-field asymptotics for heat transfer between two identical Chern insulators (\ref{analytics-CI}) and two identical graphene sheets (\ref{analytics-graphene}), confirming the $1/L$ scaling seen in the full numerical computation. The similarity between these expressions points towards a unique asymptotic scaling of
$1/L$  for the near-field radiative heat transfer for any 2D materials, valid only within local optics. Indeed, it is possible to show that for any two 2D materials with conductivity tensors
independent of the transverse momentum ${\bf K}$ (no spatial dispersion included) and possibly temperature-dependent, i.e. $\matsig^{(1)}(\omega;T_1)$ and $\matsig^{(2)}(\omega;T_2)$, held at temperatures $T_1$ and $T_2$ and separated by a distance $L$, the short-distance asymptotics of the radiative heat transfer  between them can be cast in the form 
\begin{equation}
\langle S_z \rangle= \frac{\hbar}{8 \pi^3 L} \int_0^{\infty} d\omega \omega^2 [n_B(\omega,T_1) - n_B(\omega,T_2)] f(\omega;T_1,T_2) 
\end{equation}
with
\begin{eqnarray}
&& f(\omega;T_1,T_2) = 
\frac{\sigma^{(1)}_{xx,R} \; \sigma^{(2)}_{xx,R}}
{\sigma^{(2)}_{xx,R}  \left| \sigma^{(1)}_{xx} \right|^2 + \sigma^{(1)}_{xx,R}  \left|\sigma^{(2)}_{xx}\right|^2 } 
\left[
\frac{\pi}{2} + \tan^{-1} \left(
\frac{ \sigma^{(1)}_{xx,I}  \left| \sigma^{(2)}_{xx} \right|^2 + \sigma^{(2)}_{xx,I}  \left| \sigma^{(1)}_{xx}\right|^2}
{\sigma^{(1)}_{xx,R}  \left| \sigma^{(2)}_{xx}\right|^2 + \sigma^{(2)}_{xx,R}  \left| \sigma^{(1)}_{xx}\right|^2 }
\right) \right] \nonumber
\end{eqnarray}
where $\left|\sigma^{(i)}_{xx}\right|^2 =  [ \sigma^{(i)}_{xx,R}]^2 +   [ \sigma^{(i)}_{xx,I}]^2$. For notational simplicity, we have omitted the $\omega$ and $T$ dependency of the conductivities. Thus, within the local optics approximation, the $1/L$ short-distance asymptotics for radiative heat transfer is generic for any 2D materials, with only the longitudinal conductivity $\sigma_{xx}$ contributing to the result. 

As concluding remarks, it is worth stressing the limitations of our work. As mentioned in the Introduction, the unphysical divergence of the radiative heat transfer in the near-field is due to the use of the local approximation for the optical response. It is well known, both in 3D metals  \cite{Chapuis2008} and dielectrics \cite{Singer2015}, that the inclusion of spatial dispersion results in a modification of the scaling law predicted by local optics, leading to a saturation effect at very short separations, the precise distance where this happens depends on the metallic or dielectric nature of the material. Similarly, we expect that the $1/L$ scaling law for 2D materials derived in this work via the local optics approximation will also break down when spatial dispersion is included. Nonlocal optical effects in novel 2D Dirac materials, e.g. Chern insulators, have not been thoroughly studied, and the quantification of their effect on the near-field radiative heat transfer is beyond the scope this paper. For this reason, it is not possible for us at the moment of writing to ascertain the precise regime of validity of the scaling law $1/L$  for 2D Chern insulators (Figs. 2). For the case of graphene sheets studied in Fig. 4, the effect of spatial dispersion should be similar to the case of 3D metals \cite{Chapuis2008}, since the chemical potential $\epsilon_F$ is much higher than the Dirac point and the system is basically metallic. Therefore, we expect the $1/L$ scaling law in Fig. 4 to remain valid at the shortest distances shown in that figure, nonlocal effects kicking in at distances smaller than 0.1 nm. 
Future work will need to include a complete treatment of spatial dispersion in the optical response of 2D Dirac materials and the precise evaluation of the transition distance where those effects start to become relevant for near-field radiative heat transfer. The knowledge gained for the optical response of 2D Dirac materials beyond the local optics approximation would also allow to assess the regime where spatial dispersion effects
are important in other interaction phenomena between 2D Dirac materials, such as quantum friction or Coulomb drag.

\paragraph*{Acknowledgments}

We are grateful to Age Biehs for insightful discussions. Work at Los Alamos National Laboratory was carried out under the auspicies of
the NNSA of the U.S. DOE under Award number DEAC52-06NA25396.
The research leading to these results has received funding from the People
Programme (Marie Curie Actions) of the European Union's Seventh
Framework Programme (FP7/2007-2013) under REA grant agreement nº 302005.


\end{document}